\documentstyle[epsfig,sprocl]{article}

\input{psfig.sty}

\def\be{\begin{equation}}
\def\ee{\end{equation}}
\def\bea{\begin{eqnarray}}
\def\eea{\end{eqnarray}}

\begin{document}

\title{SHELL MODEL DESCRIPTION OF $^{158}$Gd}

\author{G. POPA ,
J. P. DRAAYER  }

\address{Department of Physics and Astronomy, Louisiana State University,\\
Baton Rouge, LA 70803, USA\\
E-mail: gabriela@phys.lsu.edu, draayer@lsu.edu}

\author{ J. G. HIRSCH  }

\address{Instituto de Ciencias Nucleares, Universidad Nacional Aut\'onoma
         de M\'exico,\\ Apartado Postal 70-543 M\'exico 04510 DF, M\'exico \\
E-mail: hirsch@nuclecu.unam.mx}


\maketitle
\abstracts{The pseudo-SU(3) model is used to describe the low-energy
spectra as well as $E2$ and $M1$ transition strengths in $^{158}$Gd. The
Hamiltonian includes spherical single-particle energies, the
quadrupole-quadrupole and proton and neutron pairing interactions, plus
four rotor-like terms. The parameters of the Hamiltonian were fixed by
systematics with the rotor-like terms determined through a least-squares
analysis. The basis states are built as linear combinations of SU(3)
states which are the direct product of SU(3) proton and neutron states
with pseudo-spin zero. The calculated results compare favorably with
the available experimental data, which demonstrates the ability of the
model to describe such nuclei.}

\section{Introduction} Recently the pseudo-SU(3) model was used to
successfully describe three low-lying bands in well-deformed heavy
nuclei.~\cite{Beu00} A reasonable reproduction of the fragmentation of
the $M1$ strength was also obtained. In these applications the parameters
of the interactions were determined through a least-squares fit to
energy levels below approximately $2~MeV$. In the present work the
quadrupole-quadrupole ($\tilde Q\cdot\tilde Q$) and pairing  interaction
($H_{P}^{\pi,\nu}$) strengths were fixed by systematics while  the
interaction strengths of the other terms included in the Hamiltonian were
allowed to vary to give an overall best fit to the data. As a result of
this analysis, a consistent set of parameters has been identified. Using
this parameter set in a quadrupole-quadrupole driven truncated model
space, excellent agreement with the experimental data is obtain for
$^{158}$Gd.  The theory gives correct values for the  four lowest energy
bands, the $E2$ transition probabilities, the sumrule for $M1$
transitions from the ground state, the correct positions of the $1^+$
energies, and a reasonable reproduction of the fragmentation of the $M1$
strength.

\begin{table}[t]
\caption{ Interaction strengths used in the Hamiltonian (1).}
\begin{center}
\footnotesize
\begin{tabular}{|ccccccc|}
\hline
  $\chi$ & $G_\pi$ &  $G_\nu$ &
$a$ & $b$ & $a_{sym}$ & $a_3$  \\
\hline
 0.0077 & 0.133 & 0.108 & -0.003 & 0.23 & 0.00154 &0.0000760\\
\hline
\end{tabular}
\label{param}
\end{center}
\end{table}

\section{Model Hamiltonian}
The Hamiltonian used in the present study consists of  the following terms:
\begin{eqnarray}
     H = \sum_{\sigma = \pi,\nu}(H_{sp}^{\sigma}  + G_\sigma H_P^{\sigma})
- \frac{1}{2}~\chi
\tilde Q\cdot\tilde Q
+ a J^2 + b K_J^2 + a_3\tilde C_3 + a_{sym}\tilde C_2, ~\label{eq:ham}
\end{eqnarray}
\noindent where the last four  preserve the pseudo-SU(3) symmetry,
$\tilde C_2 $ and $\tilde C_3$ being the second and third order Casimir
invariants of SU(3). The term proportional to $J^2$ represents a
small correction to the moment of inertia, $K_J^2$ breaks the SU(3)
degeneracy of the different $K$ bands within an SU(3) irreducible
representation (irrep), $\tilde C_3$ sets the position
of the $0^+$ energies relative to one another, and the last term, which is
proportional to $\tilde C_2 $, distinguishes between the A and B$_\alpha$
($\alpha = 1, 2, 3$) type internal symmetries, pushing the $1^+$ energies
which are bandheads of B$_\alpha$-type structures up relative to the
A-type symmetries.~\cite{Dra8890}

Basis states are built by strong coupling proton and neutron SU(3) irreps
and eigenstates are a linear combination of these. The most important
configurations are those with highest spatial symmetry, indeed, only
configurations with pseudo-spin equal to zero were included for the
even-even nuclei considered in this study. The results of previous
shell-model calculations, in either a standard full pf-shell
configuration space~\cite{zu95} or in an SU(3) basis~\cite{va98}, show
that the Hilbert space can be truncated to those irreps that are favored
by the quadrupole-quadrupole interaction. Based on these results, from
the set of all allowed pseudo-SU(3) irreps only 18 with the largest
values for   $C_2 = 1/4~Q \cdot Q +3/4 ~L^2$, were used in the $^{158}$Gd
calculations reported here.

\begin{figure}[t]
\footnotesize
\begin{center}
\psfig{figure=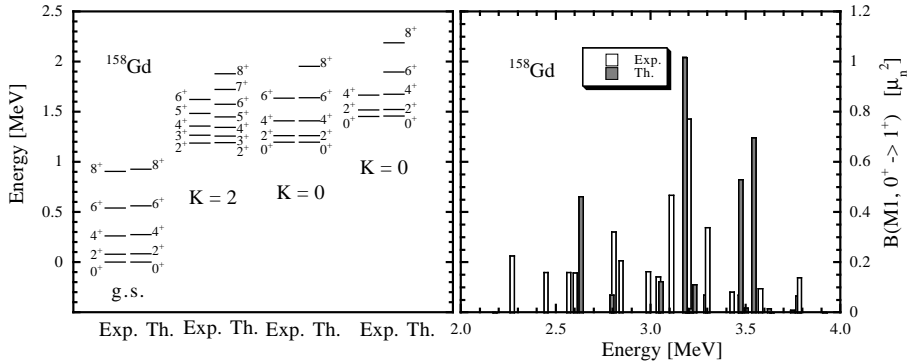,width=4.7in}
\end{center}
\label {fig:eng1}
\caption {Energy spectra of $^{158}$Gd obtained using Hamiltonian 1 and
parameters given in Table 1. The right-hand side of the figure gives
the theoretical and experimental $M1$ transition strength spectra of
$^{158}$Gd. Note that the transitions are clustered and   some of the
clusters appear to be more fragmented than others.}
\end{figure}

\section{Results} The pairing ($G_\pi$ and $G_\nu$) and
quadrupole-quadrupole ($\chi$) interaction strengths were taken from
systematics: $G_\pi = 21/A$, $G_\nu  =17/A$, and $\chi=35~A^{-5/3}$. The
dependence of the spectra on the strengths of the other terms in the
Hamiltonian was analyzed with best fit values given in
Table \ref{param}. To obtain the correct excitation energies of the
second and third $0^+$ states in the chosen model space, the
single-particle strengths needed to be reduced by a factor of four from
the so-called realistic values.~\cite{Rin79}
\begin{table}[t]
\caption{B(E2) transition probabilities in $^{158}$Gd.}
\vspace{0.2cm}
\begin{center}
\footnotesize
\begin{tabular}{|c|cc|c|cc|}
\hline
$J_i \rightarrow J_j$ & $B(E2)_{EXP}[e^2b^2]$ & $Th.[e^2b^2]$ &
$J_i \rightarrow J_j$ & $B(E2)_{EXP}[e^2b^2]$ &$Th.[e^2b^2]$ \\
\hline
$0_1 \rightarrow 2_1$ & 5.023 & 5.031 & $2_2 \rightarrow 4_1$ &
0.00137 & 0.00604\\
$2_1 \rightarrow 4_1$ & 2.639 & 2.590 & $2_2 \rightarrow 2_1$ &
0.0299  & 0.0832  \\
$4_1 \rightarrow 6_1$ & -     & 2.268 & $2_2 \rightarrow 0_1$ &
0.0177  & 0.2395  \\
$6_1 \rightarrow 8_1$ & 2.123 & 2.121 & $2_4 \rightarrow 4_1$ &
0.00705 & 0.00128\\
                         &       &       & $2_4 \rightarrow 0_1$ &
0.00157 & 0.000196 \\
\hline
\end{tabular}
\label{trans}
\end{center}
\end{table}

  Figure 1 shows the energy levels for $^{158}$Gd which are    in
excellent agreement with the experimental values.~\cite{nndc} Results for
selected (intraband and interband) transitions along with their
experimental counterparts are given in Table 2. The energy levels
belonging to one band exhibit almost the same SU(3) structure. It was also
found that for the lowest four bands only five SU(3) irreps contribute
more than $2\%$ to the eigenstates. The SU(3) content of the ground state
band is given in Table
\ref{coef}. Notice that the SU(3) irreps are only of the even-even type.
The same basic structure was found for members of the $K=2$ band. The
second and third $K=0$ bands exhibit different SU(3) content. For example,
in  Table \ref{coef}, the
second $K = 0$ band shows an approximately equal mixture of five
even-even SU(3) irreps. This
mixture supports the hypothesis that the second $K=0$ band is not
dominated by a single shape
as different SU(3) irreps correspond to different intrinsic shapes.

\begin{table}[t]
\caption{SU(3) content of calculated eigenstates for members of the
ground state and $K = 0^+_2$ bands in $^{158}$Gd. Only irreps that
contribute more than 2$\%$ to the eigenstates are shown.}
\vspace{0.2cm}
\begin{center}
\footnotesize
\begin{tabular}{|c|ccc|ccccc|}
\hline
band & ($\lambda,\mu$) &  ($\lambda_\pi,\mu_\pi$) &
($\lambda_\nu,\mu_\nu$) &
$0^+$ & $2^+$ & $4^+$ & $6^+$ & $8^+$ \\
\hline
$K=0^+_1$ &( 28,  8) & ( 10,  4) & ( 18,  4) & 78.1 & 78.8 & 80.5 &
83.0 & 85.8\\
(ground  &( 30,  4) & ( 10,  4) & ( 18,  4) & 4.3 & 4.0 & 3.4 & 2.4 & -  \\
state)& ( 30,  4) & ( 10,  4) & ( 20,  0) & 5.6 & 5.4 & 5.0 & 4.3 & 3.4 \\
& ( 30,  4) & ( 12,  0) & ( 18,  4) & 8.5 & 8.3 & 7.8 & 7.1 & 6.2 \\
&( 32,  0) & ( 10,  4) & ( 18,  4) & 2.5  & 2.4  & 2.2  & -  & - \\
\hline
$K=0^+_2$
&( 28,  8) & ( 10,  4) & ( 18,  4) & 19.2 & 18.3 & 16.0 & 17.5 & 21.6 \\
&( 30,  4) & ( 10,  4) & ( 18,  4) & 18.6 & 20.1 & 23.3 & 27.0 & 30.1 \\
&( 30,  4) & ( 10,  4) & ( 20,  0) & 23.7 & 24.7 & 26.2 & 26.2 & 23.5 \\
&( 30,  4) & ( 12,  0) & ( 18,  4) & 23.4 & 22.0 & 19.4 & 16.5 & 13.7 \\
&( 32,  0) & ( 12,  0) & ( 20,  0) & 15.1 & 14.8 & 13.9 & 12.3 & 10.1 \\
\hline
\end{tabular}
\label{coef}
\end{center}
\end{table}
The right-most spectrum given in Figure 1 shows the theoretical and
experimental M1 transition spectrum for $^{158}$Gd.~\cite{kn96}
The calculated $1^+$ energies are in the correct energy interval. The
experimental and theoretical M1 transitions peak at the same excitation
energies. Similar results were found for $^{156}$Gd and
$^{160}$Gd.~\cite{pop00} More
calculations of this type are needed to see if the pseudo-SU(3) model
can offer a consistent
explanation of the low-lying energy band structure and M1 properties
of rare earth nuclei.

\section*{Acknowledgments}
Supported in part by Conacyt (Mexico) and the National Science Foundation
(U.S.) through a Cooperative Research grant (INT-9500474).
National Science Foundation support from a regular grant (PHY-9970769)
and a Cooperative Agreement (EPS-9720652), that includes matching from the
Louisiana Board of Regents Support Fund, is also acknowledged. GP thanks
the ``Charles E. Coates Memorial Award'' for travel support.

\end{document}